\documentstyle[11pt,newpasp,twoside,psfig]{article}
\index{summary}
\index{instructions}
\index{template}

\def\edcomment#1{\iffalse\marginpar{\raggedright\sl#1\/}\else\relax\fi}
\reversemarginpar

\begin{document}

\title{A Sensitive Targeted Search Campaign at Parkes to Find Young Radio
Pulsars at 20 cm}

\author{F. Crawford} 
\affil{Lockheed Martin Management and Data Systems, P.O. Box 8048,
Philadelphia, PA 19101, USA}

\author{M. J. Pivovaroff}
\affil{Space Sciences Laboratory, University of California, Berkeley,
260 SSL \#7450, Berkeley, CA 94720-7450, USA}

\author{V. M. Kaspi}
\affil{Department of Physics, McGill University, 3600 University Street, 
Montreal, Quebec, H3A 2T8, Canada}

\author{R. N. Manchester}
\affil{ATNF, CSIRO, P.O. Box 76, Epping, NSW 1710, Australia}

\begin{abstract}
We describe a sensitive targeted search campaign at Parkes to find
faint young radio pulsars using the center beam of the 20-cm multibeam
receiver.  The high frequency of the receiver mitigates scattering
effects which can dominate profiles of distant pulsars at low
frequencies and prevent detection.  The targets included several
composite supernova remnants (SNRs), two fast X-ray pulsars,
long-period anomalous X-ray pulsars, and a soft gamma repeater.  The
sensitivity limits in this search are significantly better than those
previously reported for these objects.  We have discovered three new
radio pulsars, two of which are not associated with the target
objects. The third pulsar, PSR J1713$-$3949, is possibly associated
with SNR G347.3$-$0.5; the estimated ages and distances of the two
objects are consistent, but a more significant positional coincidence
must be established with the central X-ray source in the SNR through
pulsar timing to confirm or refute the association.
\end{abstract}

\section{Introduction}

The highly successful Parkes Multibeam Pulsar Survey (PM Survey) and
other recent surveys have used the 20-cm multibeam receiver on the
Parkes telescope for pulsar searches (e.g., Manchester et al.\ 2001;
Crawford et al.\ 2001). Apart from the advantages of the multiple
beams and good raw sensitivity of this receiver, deleterious pulse
scattering and dispersion smearing effects are greatly reduced at the
high observing frequency. This instrument is therefore good for
finding fast, distant pulsars which may have been missed in previous
low-frequency surveys.

\section{Search Observations}

We have used the multibeam receiver and the Parkes telescope to search
young objects for radio pulsars. The list of objects is shown in Table
1 and includes supernova remnants (SNRs), anomalous X-ray pulsars
(AXPs), fast X-ray pulsars, and a soft gamma repeater (SGR).  In the
search, we used the center beam of the multibeam receiver at a center
frequency of 1374 MHz. Integration times of 16800 s were typically
used with 0.25 ms sampling and 288 MHz of bandwidth split into 96
channels.  During processing, data were dedispersed at trial
dispersion measures (DMs) ranging from 0 to 3000 pc cm$^{-3}$,
frequency channels were summed, and each resulting time series was
filtered then Fourier transformed to get an amplitude spectrum for
each DM trial.  Candidate periods were identified in the spectra, and
the original data were dedispersed and folded at these periods.  The
flux sensitivity limits of the searches in all cases are $S_{\rm min}
\la 0.3$ mJy at 1374 MHz (see Table 1). These limits were estimated
using the survey sensitivity modeling described in Crawford (2000) and
Manchester et al.\ (2001), and they took into account estimates of DM,
scattering, and sky temperature, as well as instrumental effects. For
those objects for which no {\it a priori} period was known, $P = 50$
ms was assumed in the sensitivity calculations. The flux sensitivity
limits in all cases are better than those previously reported for
these objects. However, they depend on assumptions about pulsed duty
cycle, period, DM, and scattering effects.

\section{Search Results}

We obtained upper limits at 1374 MHz for most objects (Table 1), and
discovered three new radio pulsars in the search. These discoveries
are presented in Table 2 and are described below.

\begin{table} 
\caption{Target Objects Searched.} 
\begin{tabular}{lccccc} 
\tableline
Target Object & Period & Distance & DM             & $S_{\rm min}^{1374}$ & Result/PSR \\
       & (s)    & (kpc)    & (pc cm$^{-3}$) & (mJy)                &        \\ 
\tableline
SNR G0.9+0.1        & ?   & 10 $\pm$ 3       & $\sim$ 775  & 0.18 & J1747$-$2802   \\
SNR G21.5$-$0.9     & ?   & 5.5 $\pm$ 0.5    & $\sim$ 360  & 0.16 & $-$        \\
SNR G266.2$-$01.2   & ?   & $\la$ 1          & $\la$ 85    & 0.11 & $-$       \\
SNR G327.1$-$1.1    & ?   & 6.5 $\pm$ 1      & $\sim$ 330  & 0.07 & $-$       \\
SNR G347.3$-$0.5    & ?   & 6 $\pm$ 1        & $\sim$ 450  & 0.06 & J1713$-$3949   \\
A0538$-$66          & 0.069  & $\sim$ 50     & $\sim$ 100  & 0.04 & $-$       \\
AX J0043$-$737      & 0.087  & $\sim$ 57     & $\sim$ 100  & 0.04 & $-$       \\
1E 1048$-$5937  & 6.456  & $\sim$ 10?    & $\sim$ 500? & 0.13 & $-$       \\
AX J1845$-$0258 & 6.971  & $\sim$ 10?    & $\sim$ 960? & 0.13 & J1844$-$0256   \\
1E 1841$-$045   & 11.774 & $\sim$ 7      & $\sim$ 560  & 0.25 & $-$      \\
RXS J170849$-$400910 & 11 & $\sim$ 10?    & $\sim$ 800? & 0.22 & $-$       \\
SGR 0526$-$66       & 8      & $\sim$ 50     & $\sim$ 100  & 0.14 & $-$       \\
\tableline
\tableline
\end{tabular} 
\end{table} 

\begin{table} 
\caption{Newly Discovered Pulsars.}
\begin{tabular}{lccc} 
\tableline
PSR & J1713$-$3949 & J1747$-$2802 & J1844$-$0256 \\
\tableline 
$P$ (ms)                        & 392             & 2780           & 273 \\
$\tau_{c} \equiv P/2\dot{P} $   & $\sim$ 100 kyr  & $\sim$ 20 Myr  & $\sim$ 300 kyr \\
DM (pc cm$^{-3}$)               & 337 $\pm$ 3     & 795 $\pm$ 10   & 820 $\pm$ 3 \\
Distance (kpc)                  & 5.0 $\pm$ 0.2   & 11.4 $\pm$ 0.1 & 8.8 $\pm$ 0.5 \\
S/N$_{\rm detection}$           & 33.4            & 10.7           & 22.9 \\
Target Object                   & SNR G347.3$-$0.5& SNR G0.9+0.1   & AX J1845$-$0258 \\
Association?                    & possible        & no             & no \\
\tableline
\tableline
\end{tabular} 
\end{table} 

\subsection{PSR J1713$-$3949} 

PSR J1713$-$3949 has a period $P$ = 392 ms and was discovered while
searching the X-ray point source 1WGAJ1713.4$-$3949 at the center of
SNR G347.3$-$0.5 (see Figure 1). This central source is believed to be
a neutron star associated with the SNR (Slane et al.\ 1999). The
pulsar's DM of 337 pc cm$^{-3}$ indicates a distance of $\sim 5$ kpc
using the Taylor \& Cordes (1993) DM-distance model. This is
consistent with the estimated SNR distance of $6 \pm 1$ kpc (Slane et
al.\ 1999). A measured period derivative obtained from observations of
the pulsar taken a year apart indicates a characteristic age of
$\tau_{c} \equiv P/2\dot{P} \sim 100$ kyr. This value is larger than
the SNR age of $\tau_{\rm SNR} \la 40$ kyr suggested by Slane et al.\
(1999). However, these ages could still be consistent given the
uncertainties in both of the age estimates and the possibility that
$\tau_{c}$ might not reflect the true pulsar age.

The positional coincidence of PSR J1713$-$3949 with 1WGAJ1713.4$-$3949
also suggests an association with the SNR, but the uncertainty in the
pulsar's position is large ($\sim 7\arcmin$, the discovery beam
radius). A timing solution for the pulsar will provide arcsecond
positional accuracy for the pulsar as well as a more accurate age
estimate. This will enable us to establish a more significant
positional coincidence with 1WGAJ1713.4$-$3949 which can directly
confirm or refute an association between PSR J1713$-$3949 and SNR
G347.3$-$0.5.

\begin{figure}[t]
\centerline{\psfig{figure=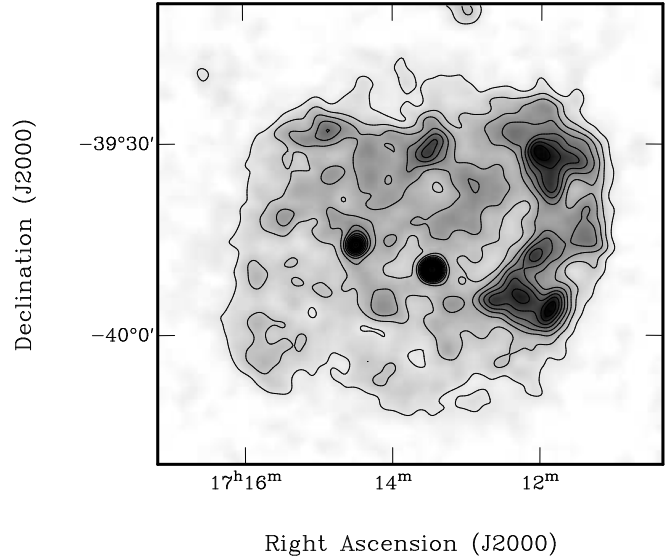,width=3.5in}}
\caption{X-ray image of SNR G347.3$-$0.5 (Figure 1 of Slane et
al. 1999). The discovery beam of PSR J1713$-$3949 was centered on the
X-ray point source 1WGAJ1713.4$-$3949 (dark circular knot) at the
center of the SNR, which is believed to be a neutron star.}
\end{figure}

\subsection{PSR J1747$-$2802 and PSR J1844$-$0256} 

PSR J1747$-$2802 is coincident with SNR G0.9+0.1 and has a period $P$
= 2780 ms and DM = 795 pc cm$^{-3}$ with a large scattering tail
($\tau_{\rm scat} \sim$ 100 ms). This pulsar was also discovered in
the PM Survey, and timing results are reported in detail by Morris et
al.\ (2001). The pulsar is old ($\tau_{c} \sim 20$ Myr) and is clearly
not associated with SNR G0.9+0.1 ($\tau_{\rm SNR} \sim$ 1-7 kyr).

PSR J1844$-$0256 was discovered in a search of AXP AX
J1845$-$0258. Its $P$ = 273 ms and DM = 820 pc cm$^{-3}$ with a large
scattering tail ($\tau_{\rm scat} \sim$ 40 ms).  The period derivative
obtained from observations taken a year apart indicates that the
pulsar is much older ($\tau_{c} \sim 300$ kyr) than the target AXP
($\tau_{\rm AXP} \la 10$ kyr), which is associated with SNR G29.6+0.1
(Gaensler, Gotthelf, \& Vasisht 1999). The pulsar's age and observed
period indicate that there is no association with AX J1845$-$0258.

\section{Conclusions} 

We have used the 20-cm multibeam receiver at Parkes to conduct
targeted pulsar searches of young objects. Our upper limits on radio
pulsations from these objects are the best yet obtained, but they
depend on assumptions about pulsed duty cycle, scattering effects, and
DM. Three new pulsars were discovered in the search; two of the
pulsars (PSRs J1747$-$2802 and J1844$-$0256) are not associated with
the target objects, while PSR J1713$-$3949 might be associated with
SNR G347.3$-$0.5. The agreement in age and distance suggests a
possible association. However, a more significant positional
coincidence must be established between PSR J1713$-$3949 and the
central X-ray source 1WGAJ1713.4$-$3949 in SNR G347.3$-$0.3 to confirm
or refute this conclusion.

\end{document}